\documentclass[twocolumn,showpacs,preprintnumbers,amsmath,amssymb]{revtex4}
\usepackage{graphicx}
\usepackage{epsfig}
\usepackage{dcolumn}
\usepackage{bm}

\begin{document}

\title{Heavy chiral bosons search at hadron colliders}

\author{Mihail V. Chizhov}
\affiliation{Centre for Space Research and Technologies, Faculty of Physics,\\
University of Sofia, 1164 Sofia, Bulgaria}


\begin{abstract}
The production of new spin-1 chiral bosons at the hadron colliders, the
Fermilab Tevatron and the CERN LHC, is considered. The masses of the chiral
bosons can be determined on the basis of experimental data of precise
low-energy experiments, which already indicate indirectly their existence. They
can explain, for example, the serious 4.5$\sigma$ discrepancy between the
measured and the predicted two pion branching ratio of the $\tau$ decay and the
sign of the 3.3$\sigma$ deviation of the muon (g-2) theoretical prediction from
the experimental value. Quantitative evaluations of the various differential
cross-sections of the chiral boson production at hadron colliders are made
using the CalcHEP package. It is noteworthy that the Tevatron data already hint
the existence of the lightest charged chiral boson with a mass around 500 GeV.
New Tevatron data and the LHC results will definitely confirm or reject this
indication. In the positive case the LHC would be able to discover all
predicted charged and neutral chiral bosons spanning in mass up to 1 TeV.
\end{abstract}


\maketitle

\section{Introduction}

The hadron colliders, due to their biggest center-of-mass energy
and their relatively compact sizes, still remain a main tool for
discoveries of very heavy particles. So in 1983 the two dedicated
UA1~\cite{UA1} and UA2~\cite{UA2} experiments discovered the
intermediate vector bosons at the CERN SPS Collider. The collider
energy, 540 GeV, just met the condition for a rough estimation of
the minimal center-of-mass energy about six times the predicted
mass of the weak bosons. The factor six corresponds to the fact
that the only part of the proton momentum is shared by the
(anti)quarks. At such energy the production cross-sections of both
$W$ and $Z$ bosons have the nanobarn level. One faces, however, a
very large background from the strong interactions.

In order to detect the production of the heavy bosons, only events with bosons
pure leptonic decays into isolated high transverse-momentum leptons and without
prominent associated jet activity have been selected. Such selection supplies
backgroundless conditions for the detection of the resonantly produced charged
$W$ bosons and more than three orders of magnitude smaller background from the
Drell--Yan dilepton production under the neutral $Z$ boson peak. At the same
time the signal from the decays of these bosons will play the role of a
background for searching of new heavy intermediate bosons, $W'$ and $Z'$, with
similar couplings to the quarks and leptons.

In any case, besides the simple manifestation of existence of the weak bosons,
one needs precise study of their properties following from the Standard Model
(SM). This task has been excellently fulfilled by the Large Electron-Positron
(LEP) storage ring at CERN and the Stanford Linear Collider (SLC) at SLAC.
Unfortunately, the masses of the $t$ quark and the undiscovered yet Higgs boson
happened to be too high to be discovered at these colliders. Nevertheless, the
precision of the electroweak measurements at the lepton colliders was so high,
that the predicted from radiative loop corrections mass of the top-quark
$m_t=180\,^{+8}_{-9}\;^{+17}_{-20}$~GeV~\cite{LEPEWG} has been found in
agreement and with comparable accuracy of its first direct measurements at the
Fermilab Tevatron by the CDF~\cite{tCDF} $m_t=176\pm 8\pm 10$~GeV and the
D0~\cite{tD0} $m_t=199\,^{+19}_{-21}\pm 22$~GeV collaborations.

In spite of the overwhelming background for the top-quark pair
production in the strong interactions at the hadron collider, the
uncertainty of the top-quark mass $m_t=170.9\pm 1.1\pm
1.5$~GeV~\cite{mt} is considerably reduced at present. Moreover,
recently, the evidence for a single top-quark
production~\cite{singleTop} through the weak interactions and
direct measurement of $|V_{tb}|$ at the Fermilab Tevatron hadron
collider became possible. Another achievement in precise
measurements at the hadron collider is the $W$-mass measurement
$m_W=80.413\pm 0.048$~GeV~\cite{CDFmW} by the CDF collaboration at
comparable with the LEP experiments accuracy, which represents the
single most precise measurement to date. All these measurements
will allow further constrain the mass of the Higgs particle, which
discovery is the main priority task of the running Tevatron and
the constructing Large Hadron Collider (LHC).

Discovery of the predicted theoretically heavy particles and
establishing of the SM without any surprises are signifying the
experimental high energy physics for the last thirty years.
Therefore, the LHC construction is connected not only with the
Higgs discovery, but with the hope to find the physics beyond the
SM. Needless to say, that the discovery of new particles and new
interactions at the hadron colliders requires a strong input of
theoretical models in order to extract the signals over huge
background.

The aim of the present paper is to point out the signature for the resonant
production of heavy spin-1 {\em chiral} bosons and their decays. It has become
proverbial (see, for example, the textbook~\cite{textbook}), that the {\em
Jacobian peak} in the transverse momentum/mass distribution is characteristic
of all two-body decays. It is not, however, the case for the decay of the new
chiral bosons~\cite{production}. To my knowledge such bosons for the first time
were introduced by Kemmer~\cite{Kemmer} and are naturally appeared in the
extended conformal supergravity theories~\cite{supergravity}.

In the next section I give the motivation of introduction of such
kind new bosons and describe their model-independent properties.
Further in section III I suggest a simple model for the extension
of the SM in order to make more definite predictions. Its
phenomenological consequences are used for explanation of the
anomalies in the precise low-energy experiments and as guiding
line for the detection and identification of such bosons at the
hadron colliders, the Fermilab Tevatron and the CERN LHC. In
section IV various distributions are produced using the CalcHEP
package~\cite{CalcHEP} and the first indications of the production
of the lightest charged chiral bosons at the Tevatron are
presented. In conclusion the prospects for the LHC physics and the
modification of the PYTHIA program~\cite{PYTHIA} for simulations
of the chiral boson production and their decays are discussed.

\section{Introduction of the chiral bosons and their model-independent
properties}

The trilinear interactions, wherein various bilinear fermion currents couple to
appropriate bosons with dimensionless coupling constants, play a key role in
the interactions of the elementary particles. They include both the gauge
couplings of the (axial-)vector fermion currents to the (axial-)vector bosons
and the Yukawa couplings of the (pseudo)scalar fermion currents to the
(pseudo)scalar bosons. Up to now only these interactions have found
phenomenological applications. For example, they have been used as effective
couplings of various mesons to quarks/baryons in the model description of the
strong nuclear interactions. On a more fundamental level of the elementary
particles, the gauge interactions and the Yukawa couplings of the Higgs bosons
to the matter constitute the base of the SM. The latter seem today a viable
solution of acquiring masses of the initially massless fermions.

In this paper I consider additional trilinear interactions, which naturally
appear in the full set of derivativeless couplings of fermion currents to
bosons and which have been missed in phenomenological applications. Let us
consider all bilinear hermitian combinations of the fermion fields. In the
relativistic case spin-$\frac{1}{2}$ fermion fields are described by the {\em
four}-component Dirac {\em bispinor} $\psi$. Therefore, there are 16
independent bilinear combinations $\bar{\psi}{\cal O}\psi$:
\begin{eqnarray}\label{bilinear}
  {\cal S}=\bar{\psi}\psi,\hspace{0.2cm}&&
  {\cal P}=i\bar{\psi}\gamma^5\psi,\nonumber\\
  {\cal V}^\mu=\bar{\psi}\gamma^\mu\psi,\hspace{0.2cm}&&
  {\cal A}^\mu=\bar{\psi}\gamma^\mu\gamma^5\psi,\nonumber\\
  &&\hspace{-1.5cm}{\cal T}^{\mu\nu}=\bar{\psi}\sigma^{\mu\nu}\psi,
\end{eqnarray}
which can couple to the $S$ scalar, $P$ pseudoscalar, $V_\mu$ vector, $A_\mu$
axial-vector and $T_{\mu\nu}$ rank-2 antisymmetric tensor fields,
correspondingly,
\begin{eqnarray}\label{ffb}
  {\cal L}_Y&=&g_S~\bar{\psi}\psi~S+ig_P~\bar{\psi}\gamma^5\psi~P
\nonumber\\
&+&g_V~\bar{\psi}\gamma^\mu\psi~V_\mu+g_A~\bar{\psi}\gamma^\mu\gamma^5\psi~A_\mu
\nonumber\\
&+&\frac{t}{2}~\bar{\psi}\sigma^{\mu\nu}\psi~T_{\mu\nu},
\end{eqnarray}
with dimensionless coupling constants.

All these bilinear combinations represent either spin-0 or spin-1 states in
accordance with the relation {\boldmath $\frac{1}{2}\otimes\frac{1}{2}=0\oplus
1$}. However, in contrast to the non-relativistic case, there are {\em two
different} spin-$\frac{1}{2}$ fermions, left-handed $\psi_L=$
\mbox{$\frac{1}{2}(1-\gamma^5)\psi$} and right-handed
$\psi_R=\frac{1}{2}(1+\gamma^5)\psi$, which transform under {\em inequivalent}
representations of the Lorentz group ($\frac{1}{2}$,0) and (0,$\frac{1}{2}$),
correspondingly. Therefore, the number of the independent combinations is twice
enlarged in comparison with the non-relativistic case. The new scalar
$\bar{\psi}\gamma^0\psi$ and pseudoscalar $\bar{\psi}\gamma^0\gamma^5\psi$
components are associated with the fourth components of the relativistic vector
$\bar{\psi}\gamma^\mu\psi$ and axial-vector $\bar{\psi}\gamma^\mu\gamma^5\psi$
currents and do not lead to any new physical spin-0 states. While the new
$\bar{\psi}\sigma^{0i}\psi$ vector and $\bar{\psi}\sigma^{ij}\psi$ axial-vector
spin-1 states are introduced as components of the independent second rank
antisymmetric tensor current $\bar{\psi}\sigma^{\mu\nu}\psi$.

In other words, in the relativistic case there are {\em two different} spin-1
states which transform under the {\em inequivalent} vector
($\frac{1}{2}$,$\frac{1}{2}$) and {\em chiral} (1,0)+(0,1) representations of
the Lorentz group. The former four-component representation is associated with
well-known \mbox{(axial-)}vector bosons, $A_\mu$ and $V_\mu$, while the latter
six-component representation demands introduction of a new rank-2 antisymmetric
tensor field, $T_{\mu\nu}$, which describes simultaneously three-components
vector and axial-vector bosons.

The new tensor Yukawa coupling in (\ref{ffb}) can be rewritten explicitly in its
chiral form
\begin{equation}\label{YukawaT}
  {\cal L}^T_Y=
\frac{t}{2\sqrt{2}}~\overline{\psi_L}\sigma^{\mu\nu}\psi_R~T^+_{\mu\nu}
+\frac{t}{2\sqrt{2}}~\overline{\psi_R}\sigma^{\mu\nu}\psi_L~T^-_{\mu\nu},
\end{equation}
where chiral components $T^\pm_{\mu\nu}=\frac{1}{\sqrt{2}}(T_{\mu\nu}\pm
i\tilde{T}_{\mu\nu})$ of the antisymmetric tensor field and its dual
$\tilde{T}_{\mu\nu}=\frac{1}{2}\epsilon_{\mu\nu\alpha\beta}T^{\alpha\beta}$
were introduced. These components are connected through the $C$
charge-conjugate and the $P$ parity transformations in the Minkowski space and
are (anti)selfdual tensors in the euclidean space.

In order to find the free Lagrangian for the new \mbox{rank-2}
antisymmetric tensor field it is enough to evaluate the one-loop
radiative correction into its self-energy from virtual fermion
pairs with known propagators. Since the Yukawa coupling constant
$t$ is dimensionless, the theory is formally renormalizable and
the structure of the quantum corrections should reproduce the bare
free Lagrangian. Simple calculations lead to conformally invariant
Lagrangian
\begin{equation}\label{LT0}
  {\cal L}^T_0=\frac{1}{4}~\partial_\rho T_{\mu\nu}~\partial^\rho T^{\mu\nu}
-\partial_\mu T^{\mu\rho}~\partial^\nu T_{\nu\rho}
\end{equation}
for the antisymmetric tensor field with mass dimension
one~\cite{Dobrev}. The properties of this Lagrangian have been
investigated in \cite{queer}. Here I just mention that in contrast
to the gauge bosons, the massless vector and axial-vector bosons,
which are described by the antisymmetric tensor field, have only
longitudinal physical components.

It is interesting to note that the mass term
$T_{\mu\nu}T^{\mu\nu}$ is not generated by the quantum
corrections, because it is protected by the chiral symmetry with
transformations
\begin{eqnarray}\label{G5}
\psi&\to&\exp[i\theta\gamma^5]\psi,\hspace{1.1cm}
\bar{\psi}\to\bar{\psi}\exp[i\theta\gamma^5],
\nonumber\\
T^+_{\mu\nu}&\to& \exp[-2i\theta]T^+_{\mu\nu},\hspace{0.4cm} T^-_{\mu\nu}\to
\exp[+2i\theta]T^-_{\mu\nu}.
\end{eqnarray}
In the same time the chirally invariant selfinteraction
\begin{equation}\label{self}
  {\cal L}^T_{4}=\lambda\left(T_{\mu\nu}T^{\mu\nu}T_{\alpha\beta}T^{\alpha\beta}
-4T_{\mu\nu}T^{\nu\alpha}T_{\alpha\beta}T^{\beta\mu}\right).
\end{equation}
is generated even in case of a {\em real} antisymmetric tensor
field, due to its chiral properties~\cite{Avdeev}.

The chiral symmetry can be localized $\theta\to\theta(x)$ including the
axial-vector gauge fields with transformations
\begin{equation}\label{G5l}
  A_\mu\to A_\mu+\frac{1}{g_A}~\partial_\mu\theta(x).
\end{equation}
In this case additional gauge interactions of the antisymmetric tensor field are
necessary~\cite{Avdeev}
\begin{eqnarray}\label{gaugeT}
  {\cal L}^T_A&=&2g_A\left(
\tilde{T}^{\mu\nu}\partial^\rho T_{\rho\nu}
-T^{\mu\nu}\partial^\rho\tilde{T}_{\rho\nu} \right)A_\mu \nonumber\\
&+&g^2_A\left(A_\rho A^\rho T_{\mu\nu}T^{\mu\nu} -4A_\mu A^\nu
T^{\mu\rho}T_{\nu\rho}\right).
\end{eqnarray}
These interactions lead to the negative contribution in the $\beta$-function of
the gauge coupling constant $g_A$, which exhibits {\em asymptotically free}
behavior even in the {\em abelian} case.

The chiral bosons acquire the mass analogously to the gauge bosons
through the Higgs mechanism and a {\em nonlocal} mass term is
generated~\cite{MPL}. Using the relation~\cite{TNJL}
\begin{equation}\label{T2RB}
  T_{\mu\nu}=\hat{R}_{\mu\nu}-
\frac{1}{2}~\epsilon_{\mu\nu\alpha\beta}~\hat{B}^{\alpha\beta},
\end{equation}
where $\hat{R}_{\mu\nu}=\hat{\partial}_\mu
R_\nu-\hat{\partial}_\nu R_\mu$,
$\hat{B}_{\mu\nu}=\hat{\partial}_\mu B_\nu-\hat{\partial}_\nu
B_\mu$ and $\hat{\partial}_\mu=\partial_\mu/\sqrt{-\partial^2}$,
the free Lagrangian (\ref{LT0}) for the antisymmetric tensor field
can be rewritten through the field-strength tensors of the new
vector $R_\mu=\hat{\partial}^\nu T_{\nu\mu}$ and axial-vector
$B_\mu=\hat{\partial}^\nu \tilde{T}_{\nu\mu}$ fields in a more
convenient form
\begin{equation}\label{LRB0}
  {\cal L}^T_0=-\frac{1}{4}~R^2_{\mu\nu}-\frac{1}{4}~B^2_{\mu\nu}
+\frac{M^2_T}{2}\left(R^2_\mu+B^2_\mu\right),
\end{equation}
where the mass terms take the usual {\em local} form in this representation.

The new fields automatically obey the Lorentz conditions $\partial^\mu R_\mu=0$
and $\partial^\mu B_\mu=0$ and can be considered as usual spin-1 fields. The
only difference to the gauge fields consists in the different Lorentz structure
of the trilinear interactions
\begin{equation}\label{ffRB}
  {\cal L}^T_Y=
t\hat{\partial}_\nu\left(\bar{\psi}\sigma^{\mu\nu}\psi\right)R_\mu
+it\hat{\partial}_\nu\left(\bar{\psi}\sigma^{\mu\nu}\gamma^5\psi\right)B_\mu
\end{equation}
and that both $R_\mu$ vector and $B_\mu$ axial-vector bosons are characterized
by the same coupling constant $t$ and the same mass $M_T$. Therefore, these
interactions automatically possess the chiral symmetry under transformations
\begin{equation}\label{G5RB}
\left(\begin{array}{c}
  R_\mu \\
  B_\mu
\end{array}\right)\to
\left(\begin{array}{rc}
  \cos2\theta & \sin2\theta \\
  -\sin2\theta & \cos2\theta
\end{array}\right)
\left(\begin{array}{c}
  R_\mu \\
  B_\mu
\end{array}\right).
\end{equation}
On the other hand in order to maintain the chiral symmetry of the Lagrangian
(\ref{ffb}) under the chiral transformations
\begin{equation}\label{G5SP}
\left(\begin{array}{c}
  S \\
  P
\end{array}\right)\to
\left(\begin{array}{rc}
  \cos2\theta & \sin2\theta \\
  -\sin2\theta & \cos2\theta
\end{array}\right)
\left(\begin{array}{c}
  S \\
  P
\end{array}\right),
\end{equation}
one needs to demand an equality of the coupling constants
$g_S=g_P$ and the masses $M_S=M_P$ of the spin-0 bosons.

The exchange of massive bosons with a momentum transfer $q_\mu$
defines all possible chirally invariant effective four-fermion
interactions in the Born approximation
\begin{eqnarray}\label{Leff}
  {\cal L}_{\rm eff}&=&\frac{g^2_S}{2(M^2_S-q^2)}~\bar{\psi}(1+\gamma^5)\psi
~\bar{\psi}(1-\gamma^5)\psi\nonumber\\
&&\hspace{-1cm}-\frac{g^2_V}{2(M^2_V-q^2)}\left(\bar{\psi}\gamma^\mu\psi\right)^2
-\frac{g^2_A}{2(M^2_A-q^2)}\left(\bar{\psi}\gamma^\mu\gamma^5\psi\right)^2
\nonumber\\
&&\hspace{-1.3cm}-\frac{t^2}{2(M^2_T-q^2)}
~\bar{\psi}\sigma^{\mu\rho}(1+\gamma^5)\psi \frac{q_\mu q^\nu}{q^2}
\bar{\psi}\sigma_{\nu\rho}(1-\gamma^5)\psi.
\end{eqnarray}
The specific of the effective tensor interactions consists in the peculiar
momentum dependent factor $q_\mu q^\nu/q^2$.

On one hand, this factor ensures good ultraviolet behavior and does not lead to
unitarity violation. On the other hand, it has a pole at $q^2=0$ and could
cause an infrared problem. However, in this paper only the processes with
$q^2\ne 0$ will be considered, I comment shortly its possible solution.

Due to the unusual properties of the chiral fields interactions, the Yukawa
coupling constant $t$ as well as the gauge coupling constant $g_A$ exhibits
asymptotically free ultraviolet behavior~\cite{Avdeev}, but the region of the
low-energy momentum transfers is governed by a nonperturbative physics. One
possible solutions to $1/q^2$ pole problem is the Leutwyler's
approach~\cite{selfdual} to the confinement in presence of constant selfdual
abelian background field $B$. So, nonperturbative effects may modify the
propagator in such a way $[1-\exp(-q^2/B)]/q^2$ to remove singularity at
$q^2=0$.

It is interesting to compare the angular distributions of the resonance
particle-antiparticle $s$-channel scattering via different intermediate bosons.
So the (pseudo)scalar spinless bosons do not prefer any direction in space and
lead to the isotropic distribution
\begin{equation}\label{sS}
  \frac{{\rm d}\sigma_S}{{\rm d}\cos\vartheta}\propto 1,
\end{equation}
where $\vartheta$ is the scattering angle in the center-of-mass
system between incoming and outcoming particles. While the
exchange of the (axial-)vector spin-1 bosons lead to the well
known distribution
\begin{equation}\label{sV}
  \frac{{\rm d}\sigma_V}{{\rm d}\cos\vartheta}\propto 1+\cos^2\vartheta
\end{equation}
up to the linear $\cos\vartheta$ term for $P$-parity nonconservation.

The tensor interactions also occur through the \mbox{(axial-)} vector spin-1
boson resonances, however they lead to a different from the previous case
distribution~\cite{two}
\begin{equation}\label{sT}
    \frac{{\rm d}\sigma_T}{{\rm d}\cos\vartheta}\propto \cos^2\vartheta.
\end{equation}
In this case there exists a characteristic plane, perpendicular to the beam
axis, where the emission of the final state pairs is forbidden. Needless to
say, that events with a large $p_T$ are the main signature of production of new
resonances at colliders. Therefore, the detection of chiral bosons with tensor
interactions will be a difficult task, due to the {\em dip} in the rapidity
distribution and the absence of the Jacobian {\em peak}~\cite{production}.

Another model independent feature of the tensor interactions is the absence of
the interference with the ordinary gauge interactions in the case of massless
fermions. In the real case of light fermions negligible interference with the
known gauge interactions also makes their detection in low-energy and collider
experiments difficult and allows them to escape the experimental constraints.

The bilinear fermion combinations (\ref{bilinear}) define the quantum numbers
$J^{PC}$ for corresponding boson states $S$, $P$, $V_\mu$, $A_\mu$, $R_\mu$ and
$B_\mu$ as $0^{++}$, $0^{-+}$, $1^{--}$, $1^{++}$, $1^{--}$ and $1^{+-}$,
respectively. All these quantum numbers can be assigned to the existing
quark-antiquark meson states (see Table~\ref{qqbar}).
\begin{table}[h]
  \centering
\begin{tabular}{|c|c|c|c|c|c|} \hline
& & & & &\\
  $J^{PC}$ & $0^{++}$ & $0^{-+}$ & $1^{--}$ & $1^{++}$ & $1^{+-}$ \\
\cline{1-6} & & & & &\\
  $I=0$ & $f_0$ & $\eta$, $\eta'$ & $\omega$, $\phi$, $\omega'$, $\phi'$ & $f_1$ & $h_1$\\
  $I=1$ & $a_0$ & $\pi$ & $\rho$, $\rho'$ & $a_1$ & $b_1$\\ \hline
\end{tabular}
  \caption{The quantum number assignments to the isoscalar $I=0$
and isovector $I=1$ neutral meson states.}\label{qqbar}
\end{table}

So, on one hand the new $CP$-odd chiral boson $B_\mu$ exhibits quantum numbers
$1^{+-}$, which undoubtedly should be assigned to the existing $h_1$ and $b_1$
bosons. On the other hand, the quantum numbers $1^{--}$ of the new $CP$-even
chiral boson $R_\mu$ coincide with the quantum numbers of the vector boson
$V_\mu$. Therefore, they could be mixed leading to assignment of the physical
states with the quantum numbers $1^{--}$ in pairs: $\omega-\omega'$,
$\phi-\phi'$ and $\rho-\rho'$. However, direct coupling of the {\em chirally
neutral} vector boson $V_\mu$ to the {\em chiral charged} boson $R_\mu$ is
forbidden by the chiral symmetry and can be realized only as a result of a
spontaneous symmetry breaking $\langle S\rangle_0\ne 0$ through {\em chirally
invariant} trilinear boson interaction
\begin{equation}\label{HVR}
  {\cal L}_{\chi}=g_{\chi}\left(ST_{\mu\nu}
+P\tilde{T}^{\mu\nu}\right)F^{\mu\nu},
\end{equation}
where $F_{\mu\nu}=\partial_\mu V_\nu-\partial_\nu V_\mu$.

A corresponding model has been developed in \cite{JETP}, where a
simple explanation of the dynamic properties of the spin-1 mesons
and new mass relations among them have been derived. The results
of this approach are in good agreement with the QCD sum rules, the
lattice calculations and the experimental data.
Let us imagine now, in accordance with the technicolor idea, that analogous
phenomenon may be extrapolated to the high energy physics at the Fermi scale,
where along with the gauge \mbox{spin-1} electroweak bosons $\gamma$, $Z$,
$W^\pm$ and the \mbox{spin-0} Higgs bosons $H$, additional spin-1 {\em chiral}
bosons $T$ are presented.

Since the chiral properties of the new spin-1 bosons are like
these of the Higgs bosons, they should come as doublets
$T_\mu=\left(T^+_\mu~T^0_\mu\right)$, taking into account the
$SU(2)_L\times U(1)_Y$ symmetry of the SM. However, the chirally
invariant trilinear interactions of the type (\ref{HVR}) and the
gauge interactions (\ref{gaugeT}) lead to new chiral
anomalies~\cite{MPL}. In order to cancel the anomalies, additional
doublets $U_\mu=\left(U^0_\mu~U^-_\mu\right)$ with opposite
hypercharge to $T_\mu$ are introduced. This concerns the Higgs
bosons as well, which should also be doubled
$H_1=\left(H^+_1~H^0_1\right)$ and $H_2=\left(H^0_2~H^-_2\right)$.

Therefore, in comparison with the SM, additional boson degrees of freedom are
introduced. In spin-0 sector, besides the light SM Higgs boson $h$, the neutral
$CP$-even $H$, $CP$-odd $A$ and the charged $H^\pm$ bosons should be present as
in the Minimal Supersymmetric Standard Model (MSSM). The new chiral bosons add
eight more spin-1 states: the neutral $CP$-even
$T^R_\mu=\sqrt{2}\,$Re$T^0_\mu$, $U^R_\mu=\sqrt{2}\,$Re$U^0_\mu$, $CP$-odd
$T^I_\mu=\sqrt{2}\,$Im$T^0_\mu$, $U^I_\mu=\sqrt{2}\,$Im$U^0_\mu$ and the
charged $T^\pm_\mu$, $U^\pm_\mu$ bosons. As far as the spin-0 sector is covered
by investigation in the MSSM, I concern only the spin-1 sector of the new
chiral bosons.

The trilinear chiral couplings of the new chiral bosons to the quarks
$Q^a=\left(u^a_L~d^a_L\right)$, $u^a_R$, $d^a_R$ and leptons
$L^a=\left(\nu^a_L~e^a_L\right)$, $\nu^a_R$, $e^a_R$
\begin{eqnarray}\label{Y}
  {\cal L}^T_Y=&&\hspace{-0.3cm}\left[
  t^q_{ab}\left(\bar{Q}^a\sigma^{\alpha\beta}d^b_R\right)+
  t^\ell_{ab}\left(\bar{L}^a\sigma^{\alpha\beta}e^b_R\right)
  \right]\left(
  \begin{array}{c} \hat{\partial}_\alpha T^+_\beta \\
\hat{\partial}_\alpha T^0_\beta \end{array}
  \right)
  \nonumber \\
  &&\hspace{-0.8cm}+\left[u^q_{ab}
  \left(\bar{Q}^a\sigma^{\alpha\beta}u^b_R\right)+
  u^\ell_{ab}\left(\bar{L}^a\sigma^{\alpha\beta}\nu^b_R\right)
  \right]\left(
  \begin{array}{c} \hat{\partial}_\alpha U^0_\beta \\
\hat{\partial}_\alpha U^-_\beta \end{array}
  \right)
\nonumber\\&&+{\rm h.c.},
\end{eqnarray}
are uniquely fixed by the symmetries. Here $a$, $b$ are generation
indexes and $t^q_{ab}$, $t^\ell_{ab}$, $u^q_{ab}$, $u^\ell_{ab}$
are in general arbitrary Yukawa coupling constants. In contrast to
ref.~\cite{MPL}, additional couplings to the right-handed neutrino
states, which are missing from the SM, are included for
generality. The latter could be important for the neutrino physics
beyond the SM such as leptogenesys and oscillations. However, in
the following I suggest that the right-handed neutrino states are
very heavy ($m_{\nu_R}\gg m_t$) and have decoupled at the
electroweak scale, i.e. one can put $u^\ell_{ab}=0$.

The explicit form of the interactions (\ref{Y}) gives us the possibility to
derive low-energy effective Lagrangian and to investigate the signature of the
chiral bosons productions at high energies. These topics with their
phenomenological consequences are considered in the next sections. In order to
be more definite in the predictions, some model-dependent simplifications are
made.

\section{Low-energy indications of new chiral interactions}

In this section I present phenomenological signatures of the chiral bosons
effects on the low-energy physics. At the beginning I assume that the chiral
bosons are very heavy to be directly produced at low energies, and their effect
is only possible through their virtual exchange. For this purpose the effective
Lagrangian approach is used. The possibility of the light (even massless)
chiral bosons is also viable, but in this case one deals with their unnaturally
small Yukawa coupling constants in order to avoid the experimental constraints.

The chiral tensor interactions are topologically similar to the electroweak
gauge interactions and, therefore, should contribute to all electroweak
processes. However, it is difficult to detect them experimentally on the SM
background and some guiding principle is necessary in order to distinguish them
from the known interactions. For example, neutral weak currents were detected
in the deep-inelastic electron scattering through measurements of $P$-odd
quantities. In the case of the new tensor interactions their {\em chiral}
properties have the main importance.

There are processes, like pion weak decays, where the SM transitions are
chirally suppressed and the tensor interactions can manifest themselves at full
strength. However, the matrix element of the tensor quark current $\langle
0|\bar{q}\sigma^{\mu\nu}q|\pi\rangle$ is zero by kinematic reasons, and there
is no direct contribution to the pion decay from the chiral boson exchange as
in the case of the (pseudo)scalar Higgs bosons~\cite{Shankar}. This allows the
tensor interactions to escape severe experimental constraints. Nevertheless,
they can contribute indirectly in the pion decay through the electromagnetic
radiative corrections~\cite{Voloshin} and directly in the radiative pion decay.
Namely, the experimental anomalies, which have been observed in the radiative
pion decays~\cite{Bolotov}, serve us for the construction of effective new
tensor interactions.

Let us consider the charged current transitions, where the only background for
the new interactions are the ordinary weak interactions. Searching for
deviations from the SM in the neutral current processes on the huge background
from the electromagnetic and weak interactions is a more challenging task.
However, due to very precise experiments in the determination of chirally
suppressed quantities as the anomalous magnetic moments for the electron and
muon, it is possible to obtain some evidence for new physics in the neutral
sector of the SM, too. This topic will be discussed at the end of this section.

Before proceeding with quantitative calculations it is necessary to fix the
arbitrary (in general) Yukawa coupling constants in (\ref{Y}). At present,
there is no cogent principle to do this. The simplest but, of course, not
unique solution, is to assume quark-lepton and family universality of the
tensor interactions
\begin{equation}\label{tu1}
  t^q_{ab}=t^\ell_{ab}=t\,\delta_{ab},\hspace{1.2cm}
  u^q_{ab}=u\,\delta_{ab}.
\end{equation}
This suggestion and the additional hypothesis about a dynamical generation of
kinetic terms for the bosons lead to the following relations among the new
coupling constants~\cite{production}
\begin{equation}\label{tug}
  t=\frac{\sqrt{3}}{2}\,u=g,
\end{equation}
where $g$ is $SU(2)_L$ gauge coupling constant. To convince ourselves in the
correctness of this assumption let us note that an analogous relation among
Yukawa coupling constants of the various hadron meson resonances in the quark
model has successful phenomenological applications~\cite{JETP}.

In order to obtain the effective low-energy tensor interactions in the limit of
heavy chiral bosons it is necessary to assume the pattern of the chiral
symmetry breaking. The most general mass term for the two charged bosons has
the form
\begin{equation}\label{MTU}
  {\cal M}^2=\left(\begin{array}{cc} T^+_\alpha & U^+_\alpha
  \end{array}\right) \left(
  \begin{array}{cc}
    M^2 & -\mu^2 \\
    -\mu^2 & m^2
  \end{array}\right)
  \left(\begin{array}{c}
    T^-_\alpha \\ U^-_\alpha
  \end{array}\right)
\end{equation}
with the only requirement of positivity of the determinant $\Delta=M^2
m^2-\mu^4=M^2_L M^2_H>0$ of the square mass matrix, where $M^2_L$ and $M^2_H$
are its eigenvalues corresponding to the lighter and heavier physical mass
states. Then the effective tensor quark-lepton interactions read
\begin{eqnarray} \label{eff}
{\cal L}^{\rm eff}_T=\hspace{-0.2cm}&-&\hspace{-0.2cm}\sqrt{2}f_T
G_F\,\bar{u}\sigma_{\alpha\rho}d_L\, \frac{4q^\rho q_\beta}{q^2}\,
\bar{e}\sigma^{\alpha\beta}\nu_L
\nonumber\\
&-&\hspace{-0.2cm}\sqrt{2}f'_T G_F\,\bar{u}\sigma_{\alpha\rho}d_R\, \frac{4q^\rho
q_\beta}{q^2}~ \bar{e}\sigma^{\alpha\beta}\nu_L+{\rm h.c.},
\end{eqnarray}
where
\begin{equation}\label{fT}
  f_T=\frac{2 M_W^2 \mu^2}{\sqrt{3}M^2_L M^2_H}>0,
\hspace{0.5cm}
  f'_T=\frac{M_W^2 m^2}{M^2_L M^2_H}>0
\end{equation}
are positive dimensionless coupling constants, which determine the strength of
the new tensor interactions relative to the ordinary weak interactions.

The experimental data on the radiative pion decays $\pi\to
e\nu\gamma$~\cite{Bolotov} show big ${\cal O}(10\%)$ deficit of events in the
high-$E_\gamma$/low-$E_e$ kinematic region. As a matter of fact both terms in
(\ref{eff}) with the coupling constants of the order of $10^{-2}$ could explain
the deficit just in the same region~\cite{discovery} through destructive
interference of the tensor interactions with the inner bremsstrahlung
radiation. However, owing to the electromagnetic radiative corrections, the
$P$-odd tensor quark current $q^\beta\bar{u}\sigma_{\alpha\beta}\gamma^5 d$,
which is present in both terms, leads to a generation of the pseudoscalar quark
current $\bar{u}\gamma^5 d$, to which pion decay is very sensitive. Therefore,
there is a severe constraint~\cite{Voloshin} from the pion decay on the
coupling constant $|f_T|<10^{-4}$ or $|f'_T|<10^{-4}$. In order to avoid this
constraint one can assume that the new tensor quark current conserves
$P$-parity due to some unknown symmetry principle and the only $P$-even tensor
quark current $q^\beta\bar{u}\sigma_{\alpha\beta} d$ is present in the
effective interactions (\ref{eff}). Because of parity conservation in
electromagnetic interactions, it does not contribute to pseudoscalar pion
decay. This is realized in the case of equality of the effective coupling
constants $f_T=f'_T$ or $\mu^2=\sqrt{3}m^2/2$.

Taking into account the latter relation, the diagonalization of the mass matrix
(\ref{MTU}) gives two mass states
\begin{equation}\label{MT}
  M^2_{H/L}=\frac{M^2+m^2\pm\sqrt{\left(M^2-m^2\right)^2+3m^4}}{2}.
\end{equation}
If the parameter $M^2$ is fixed, the maximum value of the lightest mass state is
reached at $m^2=M^2/2$, which defines the physical mass $M^2_L=M^2/4$. It
corresponds to an energetically favored exchange by this particle. Then, the
heavier state has a mass $M^2_H=M^2+M_L^2=5M^2_L$. Accepting the value of the
effective tensor coupling $f_T\approx 10^{-2}$, which can explain the deficit of
events in the radiative pion decay, one can evaluate the masses of the charged
chiral bosons
\begin{equation}\label{MHL}
  M_H=\sqrt{\frac{2}{f_T}}\,M_W\approx 1137~{\rm GeV},
  \hspace{0.3cm} M_L\approx 509~{\rm GeV},
\end{equation}
which will be used in the next section for quantitative estimations of their
production cross-sections.

All these relations are model-dependent and follow from the expressions
(\ref{tu1},\ref{tug}) for the Yukawa coupling constants. Accepting, that they
are of the order of the gauge coupling constants, the relative weakness of the
new tensor interactions are explained by the larger chiral boson masses in
comparison with the gauge bosons (see (\ref{fT})). It is interesting to note
that the heavier boson mass does not depend on the concrete value of the $t/u$
ratio, while the lighter boson mass is sensitive to it. So, in the symmetric
case, $t=u=g$, which has place when $u^\ell_{ab}\ne 0$,
$M_L=M_H/\sqrt{6}\approx 464$ GeV. In general the following limit on the mass
of the lightest charged boson $M_L<M_H/\sqrt{2}\approx 804$~GeV can be
obtained.

The same interactions (\ref{eff}) should inevitably contribute to the neutron
decay and in particular it should affect the $\lambda\equiv g_A/g_V$ and
$V_{ud}$ determination. However, there is no chiral suppression and the effect
of the new interactions is on per mille level. Nevertheless, the experimental
accuracy at present is already enough to alarm about the problem.

The most precise $\lambda$ determination follows from the electron asymmetry
parameter $A$ measured in the polarized neutron decay. The latest preliminary
result of the PERKEO collaboration $A=-0.1195\pm 0.0004$~\cite{PERKEO2006}
leads to a very high absolute value of $\lambda=-1.2755\pm0.0011$. Using the
PDG value for the neutron lifetime $\tau_n=885.7\pm0.8$ s~\cite{PDG}, one can
evaluate $V^n_{ud}=0.97081\pm 0.00088$. This value is 3.2$\,\sigma$ lower than
the extracted one from superallowed Fermi nuclear decays
$V^F_{ud}=0.97377\pm0.00027$~\cite{Vud} and may be explained through the
presence of the new interactions (\ref{eff}).

Their effect, considered in \cite{neutron}, leads to a negative contribution
into $A$ over the whole electron spectrum and to a wrong $\lambda$ evaluation
from experimental data. In the same time in a first approximation they do not
distort the recoil proton spectrum and do not contribute to the neutron
lifetime. Therefore, it is interesting to compare thus obtained $\lambda(A)$
with $\lambda(a)$ determined from the correlation coefficient $a$, relying on
the method based on measurements of the proton kinetic energy spectrum in the
unpolarized neutron decay. I hope the results of the new aSPECT experiment can
clarify the problem.

Assuming lepton universality of the tensor interactions (\ref{eff}) one can
investigate their effects in the muon and tau lepton decays, as well. They can
affect, for example, hadronic $\tau$ decays with the same effective coupling
constant $f_T$. By kinematic reason, they do not contribute to the single pion
decay channel $\tau\to\pi\nu$. At the same time, the two pion decay channel
$\tau\to\rho\nu\to 2\pi\nu$ through the vector $\rho$ meson should be
considerably affected on the level of ${\cal O}(10\%)$, due to their
interference with $\rho$ meson tensor quark current and very big mass of the
$\tau$ lepton~\cite{tau}. Using the CVC hypothesis~\cite{CVC} the branching
ratio for the two pion decay channel can be predicted from electromagnetic
process $e^+e^-\to\gamma^*\to\rho\to 2\pi$, where the tensor interactions
(\ref{eff}) are not operative. Indeed, the predicted value at present is
4.5$\,\sigma$ lower than the measured two pion branching ratio of the $\tau$
decay~\cite{Davier}.

The fact that the tensor interactions (\ref{eff}) with the same effective
coupling constant $f_T$ can explain simultaneously the destructive interference
in the radiative pion decay, the anomalously big asymmetry parameter $A$ in the
polarized neutron decay, the excess of two pion production in $\tau$ decay and
at the same time still can escape the other experimental constraints is a
witness for their vitality. The discrepancy between the two pion production in
the $e^+e^-$ annihilation and the $\tau$ decay causes another problem connected
with the prediction of the anomalous muon magnetic moment.

At present, due to the mentioned discrepancy, the tau data is not used for a
prediction of the hadron contribution to the anomalous muon magnetic moment,
while the $e^+e^-$ data lead to a value~\cite{Davier} by 3.3$\,\sigma$ lower
than measured one~\cite{mug2exp}
\begin{equation}\label{g2diff}
\delta a_\mu=a^{\rm exp}_\mu-a^{\rm th}_\mu =(27.5\pm 8.4)\times 10^{-10}.
\end{equation}
It is a huge discrepancy as far as the contribution of the massive SM weak
bosons is only $a^{\rm weak}_\mu=(15.4\pm 0.2)\times 10^{-10}$. Therefore,
naively one can expect a much lower contribution from heavier new bosons, which
cannot explain the difference (\ref{g2diff}).

Could the new neutral chiral bosons $T^0_\mu$ or $U^0_\mu$, which couple
anomalously to matter (\ref{Y}), explain the difference? Of course, at present
it is impossible, having in mind the completely unknown neutral sector of the
tensor interactions, to give quantitative value of the effect. However, it is
still possible to predict the sign of the difference. The idea is the
following.

The electromagnetic interactions of the photon $A_\alpha$ with the charged
leptons $\ell$ read
\begin{equation}\label{Lgamma}
  {\cal L}^\gamma_{\rm int}=e\,\bar{\ell}\gamma^\alpha\ell~A_\alpha
+a_\ell\frac{e}{2m_\ell}~\bar{\ell}\sigma^{\alpha\beta}\ell~\partial_\alpha
A_\beta,
\end{equation}
where the first term is the gauge interaction and the second one is the
effective photon coupling to the anomalous magnetic moment of the lepton. The
chiral structure of the effective anomalous photon coupling coincides with the
chiral structure of the new tensor interactions (\ref{Y})
\begin{eqnarray}\label{LT0Y}
  {\cal L}^N_Y=&&\hspace{-0.3cm}\frac{t}{\sqrt{2}}\left(
\bar{d}\sigma^{\alpha\beta}d+\bar{\ell}\sigma^{\alpha\beta}\ell \right)
\hat{\partial}_\alpha T^R_\beta
\nonumber \\
    &&\hspace{-0.7cm}+\,i\frac{t}{\sqrt{2}}\left(
    \bar{d}\sigma^{\alpha\beta}\gamma^5d+
    \bar{\ell}\sigma^{\alpha\beta}\gamma^5\ell\right)
    \hat{\partial}_\alpha T^I_\beta
\nonumber \\
    &&\hspace{-0.7cm}+\frac{u}{\sqrt{2}}\left(
    \bar{u}\sigma^{\alpha\beta}u\right)
    \hat{\partial}_\alpha U^R_\beta,
\nonumber \\
    &&\hspace{-0.7cm}+\,i\frac{u}{\sqrt{2}}\left(
    \bar{u}\sigma^{\alpha\beta}\gamma^5u\right)
    \hat{\partial}_\alpha U^I_\beta,
\end{eqnarray}
where only the $T_\alpha$ chiral bosons couple to the {\em charged} leptons.

An additional contribution to the anomalous magnetic moment of the lepton could
arise from a mixing between the photon $A_\alpha$ and the chiral boson
$T^R_\alpha$. In the chiral invariant limit there are not mixing between them.
However, it can appear as a result of spontaneous symmetry breaking. Simple
calculations of the mixing, for example, through the loops with the massive
fermions lead to the following contribution
\begin{equation}\label{mATR}
  {\cal L}^{\gamma T}_{\rm mix}=
-\kappa^{\gamma T}m^*_d \left(\partial^\alpha A^\beta-\partial^\beta
A^\alpha\right) \hat{\partial}_\alpha T^R_\beta,
\end{equation}
where
\begin{equation}\label{kAT}
  \kappa^{\gamma T}=\frac{\sqrt{2}\,e\,t}{4\pi^2}
\,\sum_{i=\ell,d}\frac{m_i}{m^*_d}\left(\ln\frac{\Lambda^2}{m^2_i}-1\right),
\end{equation}
$m^*_d$ is the effective mass of the {\em down} fermions and $\Lambda$ is the
effective ultraviolet cutoff.

In contrast to the case of the charged chiral bosons, the neutral states
$T^0_\alpha$ and $U^0_\alpha$ do not mix between each other, because they
couple to the different types, {\em up} and {\em down} fermions (\ref{LT0Y}).
The $CP$-odd states $T^I_\alpha$ and $U^I_\alpha$ also decouple from the
$CP$-even states $T^R_\alpha$ and $U^R_\alpha$ in the case of the $CP$
symmetry, due to their quantum numbers.

Therefore, the full free Lagrangian for the neutral bosons reads
\begin{eqnarray}\label{L0mix}
  {\cal L}_0&=&\frac{1}{2}\left(\begin{array}{cccc}
    A_\alpha & Z_\alpha & U^R_\alpha & T^R_\alpha
  \end{array}\right)
  {\cal K}^{\alpha\beta}_{\rm even}\left(\begin{array}{c}
    A_\beta \\
    Z_\beta \\
    U^R_\beta \\
    T^R_\beta
  \end{array}\right)
  \nonumber\\
  &+&\frac{1}{2}\left(\begin{array}{cc}
  U^I_\alpha & T^I_\alpha
  \end{array}\right)
  {\cal K}^{\alpha\beta}_{\rm odd}\left(\begin{array}{c}
    U^I_\beta \\
    T^I_\beta
  \end{array}\right),
\end{eqnarray}
where
\begin{eqnarray}\label{Kmixeven}
  &&\hspace{-1.1cm}{\cal K}^{\alpha\beta}_{\rm even}=
  \left(\frac{q^\alpha q^\beta}{q^2}-g^{\alpha\beta}\right)\times
  \nonumber\\
  &&\hspace{-1.1cm}\times\left(
\begin{array}{cccc}
  q^2 & \kappa^{\gamma Z}q^2 & \kappa^{\gamma U}m^*_u|q| & \kappa^{\gamma T}m^*_d|q| \\
  \kappa^{\gamma Z}q^2 & q^2-M^2_Z & \kappa^{ZU}m^*_u|q| & \kappa^{ZT}m^*_d|q| \\
  \kappa^{\gamma U}m^*_u|q| & \kappa^{ZU}m^*_u|q| & q^2-M^2_{U_R} & 0 \\
  \kappa^{\gamma T}m^*_d|q| & \kappa^{ZT}m^*_d|q| & 0 & q^2-M^2_{T_R} \\
\end{array}\right)
\end{eqnarray}
and
\begin{eqnarray}\label{Kmixodd}
  &&\hspace{-2.6cm}{\cal K}^{\alpha\beta}_{\rm odd}=
  \left(\frac{q^\alpha q^\beta}{q^2}-g^{\alpha\beta}\right)\times
  \nonumber\\
  &&\hspace{-0.8cm}\times\left(
\begin{array}{cc}
  q^2-M^2_{U_I} & 0 \\
  0 & q^2-M^2_{T_I} \\
\end{array}\right).
\end{eqnarray}
The mixings between the gauge bosons and the $CP$-even chiral bosons
(\ref{mATR}) are tiny since they are proportional to the light masses of the
ordinary fermions and the small coupling constants (\ref{kAT}). Therefore, the
matrix (\ref{Kmixeven}) is almost diagonal and the physical state of the
$CP$-even chiral boson ${\cal T}^R_\alpha$, for example, contains small but not
negligible admixture of the gauge bosons
\begin{equation}\label{mixTAZ}
  {\cal T}^R_\alpha\simeq T^R_\alpha-\kappa^{\gamma T}\,
\frac{m^*_d|q|}{M^2_{T_R}}\,A_\alpha-\kappa^{ZT}\,
\frac{m^*_d|q|}{M^2_{T_R}-M^2_Z}\,Z_\alpha.
\end{equation}
This leads to an additional positive contribution to the anomalous magnetic
moment of the lepton
\begin{equation}\label{aNP}
  \delta a_\ell\simeq\frac{\sqrt{2}\,t}{e}\,\kappa^{\gamma T}
\frac{m_\ell\,m^*_d}{M^2_{T_R}}>0.
\end{equation}
The sign of the contribution is in agreement with the discrepancy in the
experimental data for the muon (\ref{g2diff}).

Assuming the universality of the Yukawa tensor interactions and taking into
account the magnitude of the discrepancy (\ref{g2diff}) it is possible to predict
the additional contribution into the anomalous electron magnetic moment
\begin{equation}\label{aeNP}
  \delta a_e\simeq(13.3\pm 4.1)\times 10^{-12}
\end{equation}
from the photon mixing with the chiral boson. It is interesting to note that
this contribution is well above the non-QED contributions $a_e^{\rm
HAD}=1.671(19)\times 10^{-12}$, $a_e^{\rm EW}=0.030(01)\times
10^{-12}$~\cite{CODATA} and the experimental error $\delta a^{\rm
exp}_e=0.76\times 10^{-12}$ \cite{eg2exp}. Therefore, if it is real, it should
give essential correction
\begin{equation}\label{alphaNP}
  \delta \alpha^{-1}\simeq (15.7\pm 4.8)\times 10^{-7}
\end{equation}
to the determination of the fine structure constant from the anomalous electron
magnetic moment using the QED calculations~\cite{Gabrielse}. Unfortunately,
independent $\alpha$ determinations~\cite{Cs,Rb} have errors comparable with
the contribution (\ref{alphaNP}). Therefore, new $\alpha$ measurements are
badly needed. This will allow to pin down the problem with the anomalous muon
magnetic moment and the presence of the new tensor interactions.

The physical boson masses are not affected so much by the chiral symmetry
breaking. Moreover, the photon remains massless protected by the
gauge-invariant mixing (\ref{mATR}). The masses of the neutral chiral bosons
\begin{eqnarray}\label{MTU0}
  &&M_{T_R}\simeq M_{T_I}\simeq M\approx 1017~{\rm GeV},\nonumber\\
  &&M_{U_R}\simeq M_{U_I}\simeq m\approx 719~{\rm GeV},
\end{eqnarray}
are expressed through diagonal elements of the mixing mass matrix for the
corresponding doublets of the charged chiral bosons (\ref{MTU}) up to a
negligibly small correction of $\kappa^2$. This situation is completely
different from the case of the low-lying hadron meson vector states, where the
mixing is maximal and the physical masses differ considerably with respect to
the chirally symmetric case~\cite{TNJL}.

\section{Chiral boson production at the Fermilab Tevatron}

Up to now I discussed manifestations of the tensor interactions in low-energy
experiments as hints for the existence of the fundamental intermediate chiral
bosons. However, the crucial confirmation for their existence should come from
their direct production at the colliders with a unique signature.

From the previous considerations it follows that the mass of the lightest
chiral bosons is around 500 GeV. Since they are charged bosons, they could be
produced at the lepton colliders only in pairs or in association with other
charged boson, like $W$. The lightest neutral chiral bosons do not interact
with leptons and cannot be produced at the lepton colliders at all. Therefore,
to produce the pairs of the lightest charged chiral bosons or the heaviest
neutral chiral bosons, which mass is just two times bigger than the mass of the
lightest bosons, one needs a lepton collider with energy above 1 TeV. The
International Linear Collider (ILC) with such energy would be an ideal place to
produce and to study these particles.

In general the low-energy effective tensor interactions can be tested at the
lepton colliders as the LEP and the SLC or at the Hadron-Electron Ring
Accelerator (HERA). Unfortunately, these interactions do not interfere with the
SM $(V-A)$ interactions of the light particles and their contribution into the
cross-section is of the order of $f^2_T\sim 10^{-4}$. This is an order of
magnitude smaller even than the experimental errors $0.1\%$ at the
high-precision lepton colliders. Nevertheless, they can still affect the
observables connected with the heavy $\tau$ lepton and $b$ quark.

At present only the Tevatron hadron collider at Fermilab is powerful enough to
produce and detect with non-negligible probability at least the lightest
charged chiral bosons \cite{production}. Let us consider this possibility in
more detail. Since the case of the hypothetical $W'$ gauge boson with the SM
couplings is well known, in the following I shall compare its properties with
the new chiral boson's ones.

Due to the mass mixing (\ref{MTU}) the physical states are represented by two
orthogonal combinations ${\cal U}^\pm_m=(\sqrt{3}\,U^\pm_m+T^\pm_m)/2$ and
${\cal T}^\pm_m=(\sqrt{3}\,T^\pm_m-U^\pm_m)/2$, which correspond to light and
heavy massive particles, respectively. Their Yukawa interactions take the form
\begin{eqnarray}\label{YC}
    {\cal L}^C_Y=&&\hspace{-0.3cm}
    \frac{g}{2}\left(
    \bar{u}_a\sigma^{mn}d_{Ra}+
    \bar{\nu}_a\sigma^{mn}e_{Ra}\right)
    \left(\hat{\partial}_m{\cal U}^+_n+
    \sqrt{3}\,\hat{\partial}_m{\cal T}^+_n\right)
    \nonumber \\
    &&\hspace{-0.4cm}+\,g\left(\bar{u}_a\sigma^{mn}d_{La}\right)
    \left(\hat{\partial}_m{\cal U}^+_n-\frac{1}{\sqrt{3}}\,
    \hat{\partial}_m{\cal T}^+_n\right)+{\rm h.c.}
\end{eqnarray}
Special relations (\ref{tu1},\ref{tug}) between the Yukawa coupling constants
assure the same total decay widths into fermions
\begin{equation}\label{widthT}
  \Gamma^V_{tot}=\frac{g^2}{4\pi}M_V
\end{equation}
both for the gauge and chiral bosons with the same mass $M_V\gg m_t$. It
follows from the dynamical generation of kinetic terms for the bosons using the
condition of equality of all one-loop fermion contributions into self-energy
bosons, the imaginary part of which is proportional to the decay width for the
corresponding bosons. In the following, as a first approximation, I take into
account only the fermion decay channels and do not consider boson decays into
the known gauge and Higgs bosons.

The relations between the lepton and quark decay widths depend on the boson
type and the model
\begin{eqnarray}\label{widthV}
  &&\hspace{-0.8cm}\Gamma^{W'}_{ud}=\Gamma^{W'}_{cs}=\Gamma^{W'}_{tb}=
3\,\Gamma^{W'}_{e\nu_e}=3\,\Gamma^{W'}_{\mu\nu_\mu}=3\,\Gamma^{W'}_{\tau\nu_\tau},
\nonumber\\
  &&\hspace{-0.8cm}\Gamma^{\cal U}_{ud}=\Gamma^{\cal U}_{cs}=\Gamma^{\cal U}_{tb}=
15\,\Gamma^{\cal U}_{e\nu_e}=15\,\Gamma^{\cal U}_{\mu\nu_\mu}=15\,\Gamma^{\cal
U}_{\tau\nu_\tau},
\nonumber\\
  &&\hspace{-0.8cm}\Gamma^{\cal T}_{ud}=\Gamma^{\cal T}_{cs}=\Gamma^{\cal T}_{tb}=
\frac{13}{3}\,\Gamma^{\cal T}_{e\nu_e}=\frac{13}{3}\,\Gamma^{\cal
T}_{\mu\nu_\mu}=\frac{13}{3}\,\Gamma^{\cal T}_{\tau\nu_\tau},
\end{eqnarray}
or
\begin{eqnarray}\label{widthR}
  &&\Gamma^{W'}_{tot}=4\,\Gamma^{W'}_{ud}=12\,\Gamma^{W'}_{\ell\nu},
\nonumber\\
  &&\Gamma^{\cal U}_{tot}=\frac{16}{5}\,\Gamma^{\cal U}_{ud}=48\,\Gamma^{\cal U}_{\ell\nu},
\nonumber\\
  &&\Gamma^{\cal T}_{tot}=\frac{48}{13}\,\Gamma^{\cal T}_{ud}=16\,\Gamma^{\cal
T}_{\ell\nu}.
\end{eqnarray}
Here I accept a diagonal CKM mixing matrices for the $W'$ and the chiral
bosons.

The resonant production cross-section of the intermediated bosons at parton
level is proportional to their partial decay width into the quarks pair
\begin{eqnarray}\label{sigmaV}
  \hat{\sigma}(u\bar{d}\to V^+)
&=&\frac{4\pi^2}{3M_V}\,\Gamma^V_{ud}\,\delta(\hat{s}-M^2_V) \nonumber\\
&=&\frac{\pi g^2}{3}{\cal B}(V^+\to u\bar{d})\,\delta(\hat{s}-M^2_V),
\end{eqnarray}
where $\hat{s}=(p_u+p_{\bar{d}})^2$ is the invariant Mandelstam variable. Here
and in the following I shall denote with a hat the variables in the parton
center-of-mass system.

Neglecting the small contribution of the see quarks at the Tevatron $p\bar{p}$
collider, the total production cross-section reads
\begin{equation}\label{sTev}
    \sigma^{\rm Tev}_{V}\!=
    \frac{\pi g^2}{3s} {\cal B}(V\to u\bar{d})\!\!
    \int^1_\tau\hspace{-0.2cm} u(x,M_L)d\!\left(\frac{\tau}{x},M_L\right)
    \!\frac{{\rm d}x}{x},
\end{equation}
where $s$ is the square of the center-of-mass energy and $\tau=M^2_V/s$. From
the relations (\ref{widthR}) and (\ref{widthT}) it follows that the quark
branching ratios ${\cal B}(W'\to u \bar{d})\simeq 25.0\%$, ${\cal B}({\cal
U}\to u \bar{d})\simeq 31.3\%$ and ${\cal B}({\cal T}\to u \bar{d})\simeq
27.1\%$ do not differ very much for the different type of bosons, therefore,
the total production cross-sections of the gauge and the chiral bosons with the
same mass have approximately the same magnitudes.

\begin{figure}[th]
\epsfig{file=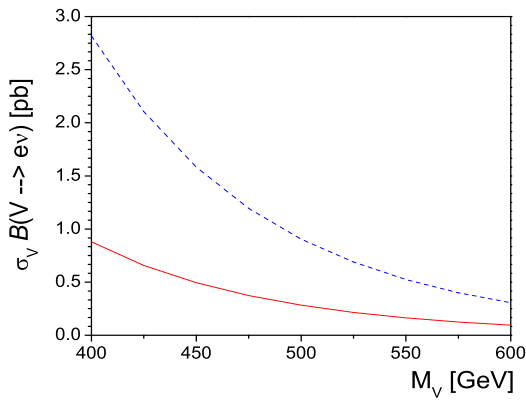,height=6.3cm,width=8cm} \caption{\label{fig:1} The
production cross-sections of the gauge $W'$ boson (dashed) and the chiral
${\cal U}$ boson (solid) as functions of their masses. }
\end{figure}

However, the cleanest way to detect the production of the heavy intermediate
bosons at the hadron colliders is to look for their practically backgroundless
decay channels into leptons. It is interesting to note that the lepton
branching of the lightest chiral boson ${\cal B}({\cal
U}\to\ell\bar{\nu})\simeq 2.1\%$ is considerably smaller than the others ${\cal
B}(W'\to\ell\bar{\nu})\simeq 8.3\%$ and ${\cal B}({\cal
T}\to\ell\bar{\nu})\simeq 6.3\%$. Therefore, its leptophobic character leads to
considerably small cross-section in the lepton channel (see Fig.\ref{fig:1}).

For these calculations the parton distribution functions CTEQ6M~\cite{CTEQ6}
and the factor $K=1$ have been used. So, for the reference mass $M_V=M_L\approx
509$~GeV (\ref{MHL}) of the lightest chiral boson, the corresponding cross
sections are
\begin{eqnarray}\label{sWpUlepton}
  &&\sigma_{W'}\times {\cal B}(W'\to\ell\bar{\nu})\approx 0.82~{\rm pb},
\nonumber\\
  &&\sigma_{\cal U}\times {\cal B}({\cal U}\to\ell\bar{\nu})\approx 0.26~{\rm pb}.
\end{eqnarray}
The latter cross-section is by factor 16/5 smaller than the former one. This fact
makes the detection of the chiral bosons in the lepton channels more difficult
than the gauge ones.

But not only this fact prevents the discovery of the new chiral bosons up to
now. Another unusual and unexpected feature of the chiral bosons, connected to
their anomalous interactions (\ref{Y}) with fermions, has place. Let us compare
the normalized angular distributions of the leptons from the decays of the
gauge $W^-$
\begin{equation}\label{NW}
    \frac{{\rm d}N_W}{{\rm d}\Omega}=\left\{
    \begin{array}{ll}
    \frac{3}{16\pi}(1\mp\cos\theta^*)^2, & \lambda=\pm1,\\
    &\\
    \frac{3}{8\pi}\sin^2\theta^*, & \lambda=0,
    \end{array}\right.
\end{equation}
and the chiral ${\cal U}^\pm$
\begin{equation}\label{NU}
    \frac{{\rm d}N_{\cal U}}{{\rm d}\Omega}=\left\{
    \begin{array}{ll}
    \frac{3}{8\pi}\sin^2\theta^*, & \lambda=\pm1,\\
    &\\
    \frac{3}{4\pi}\cos^2\theta^*, & \lambda=0,
    \end{array}\right.
\end{equation}
bosons, where $\theta^*$ is the angle of the lepton with respect to the
longitudinal axis in the boson-rest frame and $\lambda$ is the boson helicity.

For example, the left-handed quark $d$ (from the proton) interacting with the
right-handed anti-quark $\bar{u}$ (from the anti-proton) can produce $W^-$ with
spin projection on the proton beam direction $-1$. Hence, the decay leptons are
distributed as $(1+\cos\theta^*)^2$. While chiral particle production arises
from the interaction of a quark and an anti-quark with the same helicities,
that leads to zero helicity of the produced chiral boson and the
backward-forward symmetric $\cos^2\theta^*$ lepton distribution~\cite{two}.

Indeed, the cross-section for $p+\bar{p}\to {\cal U}+X\to\ell+X'$
process
\begin{equation}\label{S}
    {\rm d}\sigma=\frac{1}{3}
    \int\hspace{-0.1cm}{\rm d}x_1{\rm d}x_2\,
    u(x_1,M_L)d(x_2,M_L)\,
    {\rm d}\hat{\sigma}(\hat{s},\hat{t})
\end{equation}
is expressed through the relevant differential cross-section of the parton
subprocess $d+\bar{u}\to{\cal U}^-\to\ell+\bar{\nu}$
\begin{equation}\label{s}
    E_\ell\frac{{\rm d}^3\hat{\sigma}(\hat{s},\hat{t})}
    {{\rm d}^3 p_\ell}=
    \frac{5g^4}{(32\pi)^2}\,
    \frac{(\hat{s}+2\hat{t}\,)^2\delta(\hat{s}+\hat{t}+\hat{u})}
    {\hat{s}|\hat{s}-M_L^2+iM_L\Gamma^{\cal U}_{tot}|^2},
\end{equation}
where $\hat{s}=(p_d+p_{\bar{u}})^2$, $\hat{t}=(p_{\bar{u}}-p_\ell)^2$ and
$\hat{u}=(p_d-p_\ell)^2$ are the Mandelstam variables. In the center-of-mass
parton system the differential cross-section shows the following distribution
\begin{equation}\label{cos2}
    \frac{{\rm d}\hat{\sigma}}{{\rm d}\Omega}\propto
    (\hat{s}+2\hat{t}\,)^2\propto
    \cos^2\hat{\theta}=1-\frac{4\hat{p}^2_T}{\hat{s}}.
\end{equation}
Here $\hat{\theta}$ is the angle between the lepton and the parton
direction, which coincide with the angle $\theta^*$ in the
boson-rest frame, and $\hat{p}^2_T$ is the square of the
transverse lepton momentum.

Since the latter is invariant under longitudinal boosts along the beam direction,
the distribution (\ref{cos2}) versus $\hat{p}_T$ holds also in the lab frame
$p_T=\hat{p}_T$. Changing variables in the differential cross-section from
$\cos\hat{\theta}$ to $p^2_T$
\begin{equation}\label{Jacobian}
    \frac{{\rm d}\cos\hat{\theta}}{{\rm d}p^2_T}=
    -\frac{2}{\hat{s}}\left(\sqrt{1-\frac{4p^2_T}{\hat{s}}}\,\right)^{-1}
\end{equation}
leads to a kinematical singularity at the endpoint
$p^2_T=\hat{s}/4$, which gives the prominent Jacobian peak in the
$W$ decay distribution.

\begin{figure}[th]
\epsfig{file=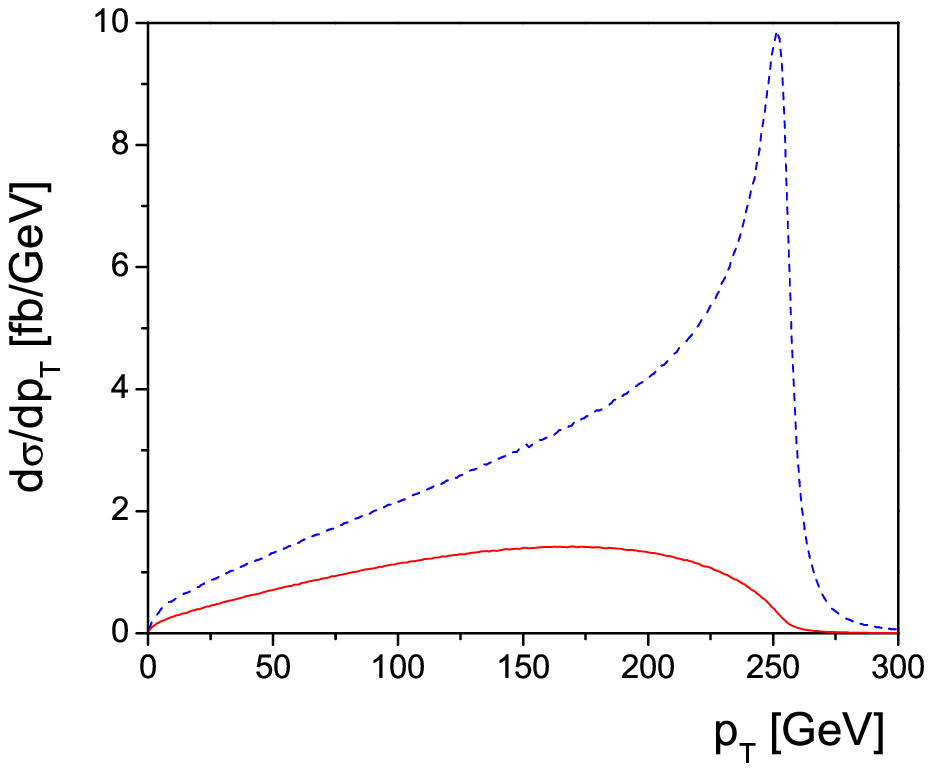,height=6.3cm,width=8cm} \caption{\label{fig:2} The
differential cross-sections for the gauge $W'$ boson (dashed) and the chiral
${\cal U}$ boson (solid) as functions of the lepton transfers momentum. }
\end{figure}

In contrast to this, the pole in the decay distribution of the chiral bosons is
cancelled out and, moreover, the distribution reaches zero at the endpoint
$p^2_T=\hat{s}/4$. The chiral boson decay distribution has a broad smooth bump
with a maximum below the kinematical endpoint, instead of a sharp Jacobian peak
(Fig.~\ref{fig:2}). In the case when the chiral boson is produced with no
transverse momentum, the transverse mass of the lepton pair is related to $p_T$
as $M_T(\ell\bar{\nu})=2p_T$ and the Jacobian peak is absent in $M_T$
distribution as well.

Therefore, the transverse momentum/mass decay distribution of the chiral bosons
differs drastically from the distribution of the gauge bosons. Even relatively
small decay width of the chiral bosons leads to a wide distribution, that
obscures their identification as resonances at hadron colliders. The form of
the decay distribution for the chiral bosons resembles the bump anomalies in
the inclusive jet $E_T$ distribution, reported by the CDF
Collaboration~\cite{CDF} many years ago. Although this problem has been solved
in the framework of the SM by changing the gluon distribution
functions~\cite{gluon}, it could be reconsidered in the light of the new form
of the decay distribution as a real physical signal from decays of different
chiral bosons, both charged and neutral.

Analysing the bumps in the jet transverse energy distribution in Fig.~1 of ref.
\cite{CDF}, we can find the endpoint of the first bump at 250 GeV and guess
about the second bump endpoint from the minimum around 350 GeV. If we assign
the first bump to the hadron decay products of the lightest charged bosons,
which exactly corresponds to the estimated mass from eq. (\ref{MHL}), the
second endpoint hints to a mass around 700~GeV, which is also in a quantitative
agreement with our estimations (\ref{MTU0}) for the mass of the lightest
leptophobic neutral boson. However, taking into account the large systematic
uncertainties in jet production, these conclusions may be premature, unless
they are confirmed in other channels.

In the following the CalcHEP~\cite{CalcHEP} package will be used for the
numeric calculations of various distributions. For these purposes I have
introduced the new chiral bosons in the package and implemented the
corresponding interactions for them. Indeed, the CalcHEP model does not allow
to introduce the interactions (\ref{Y}) directly. The problem is connected with
unusual $1/\sqrt{q^2}$ momentum dependence of the Yukawa interactions, when the
chiral bosons are described by the four-dimensional Lorentz vector $V_\alpha$
with the usual propagator $-ig_{\alpha\beta}/(q^2-M^2_V)$. It can be avoided
for the processes with the chiral particles on the mass shell $q^2=M^2_V$, when
the the function $1/\sqrt{q^2}$ can be replaced by constant factor $1/M_V$, but
not for the intermediate chiral bosons.

The solution of the problem was prompted to me by A. Pukhov, one of the authors
of the CalcHEP package. It consists in introducing of pair of massive particle
and its massless ghost in such a way that the propagator for the intermediate
states multiplied by the momentum dependent factor $1/q^2$, which comes from
the Yukawa couplings, can be represented as a difference of two propagators
\begin{equation}\label{diffP}
  \frac{1}{q^2}\,\frac{1}{q^2-M^2_V}=
  \frac{1}{M^2_V}\left(\frac{1}{q^2-M^2_V}-\frac{1}{q^2}\right)
\end{equation}
with the constant factor $1/M^2_V$. Therefore, the situation is reduced to the
previous case of the chiral boson description on the mass shell by effective
Yukawa interactions with the constant factor $1/M_V$. This dimension factor
naively indicates a problem with high-energy behaviour of scattering amplitudes
and renormalizability of the model, which are restored by the including of the
ghosts.

The distributions in the Fig.~\ref{fig:2} have been calculated without any
kinematical cuts. However, experimental detectors always have dead zones where
the particles cannot be registered. The simple examples are the
backward-forward regions along the beam direction at the colliders. The CaclHEP
package with its effective Monte Carlo integration allows simply and quickly to
calculate divers distributions with experimental cuts as a first approximations
to the detector simulations.

\begin{figure}[th]
\epsfig{file=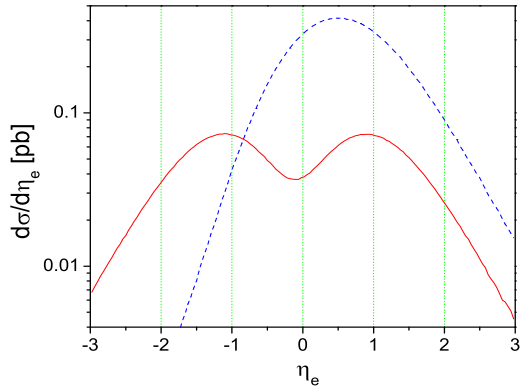,height=6.3cm,width=8cm} \caption{\label{fig:3} The
differential cross-sections for the gauge $W'$ boson (dashed) and the chiral
${\cal U}$ boson (solid) as functions of the lepton pseudorapidity. }
\end{figure}
\begin{figure}[th]
\epsfig{file=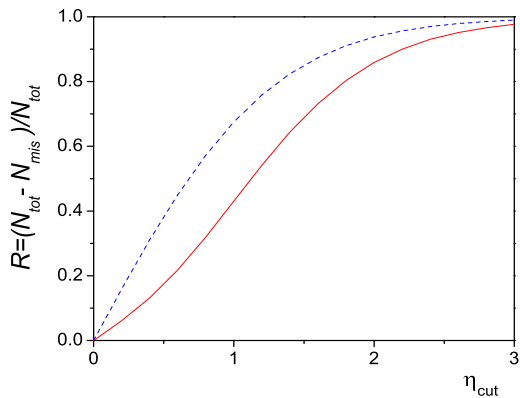,height=6.3cm,width=8cm} \caption{\label{fig:4} The
ratio of the detected with $\vert\eta_e\vert\le\eta_{cut}$ events to their
total number for the gauge $W'$ boson (dashed) and the chiral ${\cal U}$ boson
(solid). }
\end{figure}

The cuts in the backward-forward regions lead to miss in a essential part of
the events from the chiral boson decays due to the mentioned previously
specific angular distribution. So, in the Fig.~\ref{fig:3} the differential
cross-sections of the gauge $W'$ boson and the chiral ${\cal U}$ boson as
functions of the lepton pseudorapidity are shown. While the maximum of the
gauge boson distribution is centered at the small lepton pseudorapidities,
which correspond to the central part of the detector, the chiral boson
distribution has minimum in this region and its maxima are placed at the edges
of the CDF and D0 central calorimeters. Based on the fact that the major part
of the leptons stemming from the $W'$ decays are emitted in the central
detector region, both collaborations have analyzed the spectrum of the
transverse high-energy electrons only in the central electromagnetic
calorimeters $\vert\eta_e\vert\le\eta_{cut}\simeq 1$. The ratio
  $R=(N_{tot}-N_{mis})/N_{tot}$,
where $N_{mis}$ is the number of missing due to the cut events, for the $W'$
boson and the chiral ${\cal U}$ boson is shown in the Fig.~\ref{fig:4}.

\noindent As seen from Fig.~\ref{fig:4} the curve for the chiral ${\cal U}$
boson lies always under the $W'$ curve, and at $\eta_{cut}\simeq 1$ there are
around 70\% detected events in the case of the gauge $W'$ boson and only 45\%
in the case of the chiral ${\cal U}$ boson.

\begin{figure}[th]
\epsfig{file=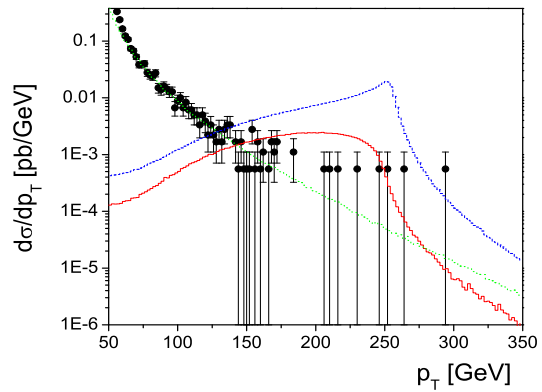,height=6.3cm,width=8cm} \caption{\label{fig:5} The
differential cross-sections for the gauge $W$ (dotted), $W'$ (dashed) and the
chiral ${\cal U}$ (solid) bosons as a function of the electron transverse
momentum versus the D0 data. }
\end{figure}

If the new chiral boson production takes place with subsequent decay into the
lepton and its antineutrino, an eventual excess should be watched in the region
$350$~GeV $<M_T<500$ GeV, where the background from the tail of the $W$ decays
is considerably small. Indeed, such an excess about 2$\sigma$ has been pointed
out recently by the CDF Collaboraion~\cite{WpCDFlepton} in the same region. In
the case of the gauge $W'$ boson production an excess should be peaked around
$M_T\approx 500$~GeV or $p_T\approx 250$~GeV. The latter was unambiguously
rejected by the D0 data~\cite{WpD0lepton} with more better calorimetry than the
CDF detector (Fig.~\ref{fig:5}).
At the same time the chiral boson distribution, probably up to some common
normalized factor, is still in agreement with the D0 experimental data. The
excess should look like as a shoulder rather than a peak and the small number
of the events in this region cannot give conclusive statement about the excess.

Nevertheless, the $p\bar{p}$ colliders such as the CERN SPS and the Fermilab
Tevatron have unique possibility to measure lepton asymmetry, which cannot be
done at the powerful but $pp$ symmetric LHC machine. So, one of the crucial
confirmations of the $W$ properties at the SPS, even with a small number of
events, was revealing the backward--forward asymmetry in the angular
distribution of the lepton emission angle $\theta^*$ in the $W$-rest
frame~\cite{UA1cos}. Application of analogous analysis for the events from the
$W$ tail at the Tevatron would allow to reduce additionally the background in
the search of the gauge right-handed $W'_R$ and the chiral ${\cal U}$ bosons.

Such analysis is even more important in the search of heavy boson production
through their hadronic decay channels, where the background from the strong
interactions is overwhelming. So, the cross-sections for the $t\bar{b}$ quark
channel are
\begin{eqnarray}\label{sWpUquark}
  &&\sigma_{W'}\times {\cal B}(W'\to t\bar{b})\approx 1.91~{\rm pb},
\nonumber\\
  &&\sigma_{\cal U}\times {\cal B}({\cal U}\to t\bar{b})\approx 3.56~{\rm pb},
\end{eqnarray}
for the intermediate states with the $W'$ and ${\cal U}$ bosons,
correspondingly. The latter cross-section is 25/16 times bigger than the former
one, due to the hadrophilic character of the chiral boson. Therefore, its
detection in the hadronic channels could be even easy than the detection of the
new gauge boson.

While the light quark decay channels are swamped by multijet background, the
$t\bar{b}$ pair of the heavy $b$ quark and the short living $t$ quark with its
subsequent decay to $Wb$ pair allow to make jet $b$-tagging, where one of the
jets must have a displaced secondary vertex. Searching for the intermediate
heavy bosons in this channel has been fulfilled by both the D0 and CDF
collaborations, as for this purpose the part of the same datasets of the single
top production analyses has been used. Owing to their high masses this analysis
is even simpler than the single top production searches, because at such
energies the background is considerably reduced.

Indeed, the CDF collaboration even with approximately 1 fb$^{-1}$ of the
$p\bar{p}$ RUN II data is still unable to pin down the single top quark
production, while a slight excess in the region near $450-500$ GeV in the
invariant mass of the reconstructed $W$ and two leading jets ($M_{Wjj}$) has
been pointed out in \cite{WpCDFquark}. The excess is seen in the \mbox{2-jet}
mass distribution histogram with 40 GeV bin's width. It is also expected to
find a Jacobian peak in the transverse momentum distribution of the leading $b$
jet or reconstructed top, however, if this excess is due to the chiral boson
production, it could not be there.

The common wisdom, that a peak in the invariant mass distribution of the two
final particles must correspond to the peaks in their transverse mass
distributions $M_T$ at the same value, is not valid for the chiral bosons. On
the other hand the peaks in the invariant mass distribution come from the
Breit--Wigner propagator form, which is the same as for the gauge and chiral
bosons in the Born approximation. However, some small different corrections
should be applied to the shape of the resonance curve for the chiral bosons,
due to their different couplings to the fermions.

It is interesting to note that this excess is in some sense a confirmation of
the excess in the leptonic channel~\cite{WpCDFlepton} of the same
collaboration. Therefore, an independent result from the D0 collaboration is
very important. Their published result~\cite{WpD0quark} is based on 230
pb$^{-1}$ of integrated luminosity and does not show any excess in the
histogram with the bin's width of 50 GeV. However, it should always be taken
into account that the narrow peak could be missed due to the smearing effect of
the detector resolution or an insufficient statistic. Indeed, the right
histogram in the Fig.~3 of the conference paper \cite{WpD0quarkNote} of the
same collaboration with the bin's width of the 45 GeV reveals, nevertheless,
the weak peak in the same region of the 500 GeV. All these not statistically
significant results for the separated analyses may give a more conclusive
answer after their combining and additional investigation of the angular
distributions of the events in this region. The only difficulty could arise
from reconstructing the full kinematics due to double solution for the neutrino
longitudinal momentum.

Concluding this section I would like to note that some hints for the existence
of the lightest charged chiral boson already exist in the Tevatron data. What
about signature of the other chiral bosons? The next to the lightest chiral
boson are the two completely leptophobic neutral $CP$-even $U^R$ and $CP$-odd
$U^I$ bosons with approximately the same mass $m\approx 719$~GeV, which couple
only to $up$-type quarks. Therefore, they could be looked for in the $t\bar{t}$
decay channel. However, the small cross-section
\begin{equation}\label{sU0Tev}
  \sigma_U\times{\cal B}\left(U\to t\bar{t}\right)\approx 1.63~{\rm pb}
\end{equation}
still hinders their revealing.

The other pair of more heavy neutral chiral bosons, $CP$-even $T^R$ and
$CP$-odd $T^I$ can be seen in their dilepton decay channels. However, their
high masses $M\approx 1017$~GeV lead to completely negligible cross-section
\begin{equation}\label{sT0Tev}
  \sigma_T\times{\cal B}\left(T\to\ell\bar{\ell}\right)\approx 2.3~{\rm fb}
\end{equation}
at the Tevatron. The heaviest charged chiral boson ${\cal T}$ with mass
$M_H\approx 1137$~GeV also leads to negligible cross-section in the lepton
channel
\begin{equation}\label{sTcTev}
  \sigma_{\cal T}\times{\cal B}\left({\cal T}\to\ell\bar{\nu}\right)\approx 2.2~{\rm fb}
\end{equation}
at the Tevatron.

\section{CERN LHC prospects and conclusions}

The LHC at CERN belongs to the next generation of hadron colliders. There is no
doubt that it will be discovery machine with its 14 TeV collider energy in the
$pp$ center-of-mass. All new particles with non-negligible coupling constants
to the ordinary matter up to a mass of the order of $2-3$ TeV could be
discovered if they exist. The new chiral bosons, with coupling constants of the
order of the gauge ones and the predicted masses, comfortably fall into this
category. Nevertheless, the unusual properties of the chiral bosons, which are
still unknown to the experimentalists, require more work in simulations.

As a first step into this direction the corresponding model was introduced in
the CalcHEP package. In the agreement with the Les Houches Accord~\cite{LHA}
the package has an interface with the PYTHIA. However, the experimental
software on simulation and reconstruction (see, for example, \cite{WpCMS})
often has itself an interface with the event generator programs like PYTHIA.
Therefore, I have built the code for the chiral bosons directly into the
PYTHIA, as well. However, the right way to do this is to use user-defined
external processes machinery. In order to reach the result in a simple way and
as soon as possible I have used the fact that the interactions of the chiral
bosons are very similar to those of the $W$ and $W'$. Therefore, I have just
slightly corrected the PYTHIA code for the subprocess 142 ($W'$ production) in
the subroutines PYWIDT and PYRESD.

The subroutine PYWIDT calculates full and partial widths of resonances. As far
as the decays of the gauge and the chiral bosons are described by different
formulas in the case of their decay into two massive fermions, I have
substituted the matrix element
\begin{eqnarray}\label{MWpPYTHIA}
  |{\cal M}_{W'}|^2&=&8M^2\left\{\left(g^2_V+g^2_A\right)
\left[2-r_1-r_2-\left(r_1-r_2\right)^2\right]\right.
\nonumber\\
&&\hspace{1.2cm}+\,6\left.\left(g^2_V-g^2_A\right)\sqrt{r_1 r_2}\right\}
\end{eqnarray}
in the expression for the $W'$ width with the following expression
\begin{eqnarray}\label{MUcPYTHIA}
  |{\cal M}_{\cal U}|^2&=&8M^2\left\{\left(g^2_V+g^2_A\right)
\left[1+r_1+r_2-2\left(r_1-r_2\right)^2\right]\right.
\nonumber\\
&&\hspace{1.2cm}+\,6\left.\left(g^2_V-g^2_A\right)\sqrt{r_1 r_2}\right\}
\end{eqnarray}
for the chiral bosons. Here $r_1=(m_1/M)^2$ and $r_2=(m_2/M)^2$ are squared
ratios of the fermions masses $m_1$ and $m_2$ to the boson mass $M$.

The parameters PARU(131)/PARU(132) and PARU(133)/PARU(134) set the vector
$g_V$/axial $g_A$ couplings of the quark and lepton pairs to the heavy vector
boson, correspondingly. These is a bug in the case of $W'$ decay with non-equal
vector and axial-vector couplings constants $|g_V|\ne|g_A|$, because the second
line of eq. (\ref{MWpPYTHIA}) is not coded into the program. Corrected in this
way the subroutine should provide correct description of the $W'$ decay with
the $W$-like coupling constants PARU(131)=PARU(133)=1, PARU(132)=PARU(134)=$-1$
or the chiral ${\cal U}$ boson with following parameters
\begin{eqnarray}\label{PARU(Uc)}
  &&{\rm PARU}(131)=\frac{3}{2},\hspace{0.3cm} {\rm PARU}(132)=-\frac{1}{2},
\nonumber\\
  &&{\rm PARU}(133)=\frac{1}{2},\hspace{0.3cm} {\rm PARU}(134)=-\frac{1}{2}.
\end{eqnarray}

The subroutine PYRESD describes angular distributions of the resonances, which
are completely different in the case of the gauge and chiral bosons. So, in
accordance with the normalized distributions (\ref{NW}) and (\ref{NU}), the
distribution
\begin{equation}\label{AWpPYTHIA}
  {\rm WT}=1+{\rm ASYM}\cdot\cos\hat{\theta}+\cos^2\hat{\theta}
\end{equation}
and its maximum value WTMAX=2+ABS(ASYM) for the gauge bosons, where the
coefficient ASYM defines backward-forward asymmetry, should be replaced by
simple formula
\begin{equation}\label{AUcPYTHIA}
  {\rm WT}=4\cos^2\hat{\theta}
\end{equation}
and a corresponding maximum value WTMAX=4 for the chiral bosons.

These changes open the possibility for full detector simulations of the
production and diverse decay channels of the charged chiral bosons. The
implementation of the neutral chiral bosons within PYTHIA cannot be fulfilled
just as trivial modifications of already existing code for the $Z'$ boson,
because the code takes into account the full $\gamma^*/Z/Z'$ interference
structure and is very complicated. In the same time the chiral bosons do not
interfere with the gauge bosons and the new part of the code should be simplier
and could be easily added by a PYTHIA expert.

\begin{figure}[thb]
\epsfig{file=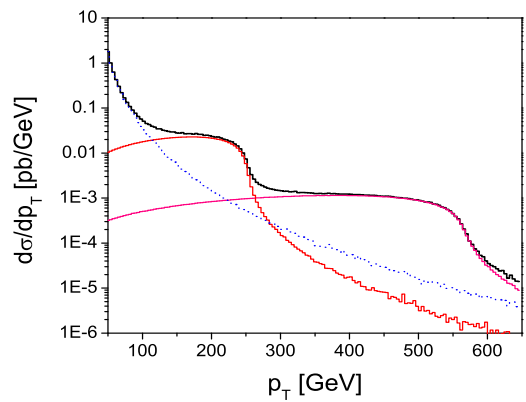,height=6.3cm,width=8cm} \caption{\label{fig:6} The
differential cross-sections for the gauge $W$ (dotted) and the chiral ${\cal
U,T}$ (solid) bosons as functions of the lepton transverse momentum. }
\end{figure}

I will continue with consideration of the golden discovery channels $pp\to
{\cal U/T}+X\to\ell\bar{\nu}+X'$ and $pp\to T^R/T^I+X\to\ell\bar{\ell}+X'$ for
the chiral bosons at the LHC using CalcHEP package. Due to the enormous
collider energy and the designed peak luminosity 10$^{34}$~cm$^{-2}$s$^{-1}$
they could be explored already at the very beginning, when the SM calibration
processes $pp\to W+X\to\ell\bar{\nu}+X'$ and $pp\to Z+X\to\ell\bar{\ell}+X'$
are studied. So, the differential cross-sections for the gauge $W$ (background)
and the chiral ${\cal U,T}$ (signal) bosons as a function of the lepton
transverse momentum are shown in the Fig.~\ref{fig:6}.

The first shoulder corresponds to the resonance production of the light charged
chiral bosons ${\cal U}^\pm$ with a differential cross-section of the order of
0.03 pb/GeV. It means, that after only an hour of data-taking at the peak
luminosity, approximately 10 events should be within each bin with transverse
lepton momentum around 200 GeV and 10 GeV bin's width. To see the second
shoulder, corresponding to the heavy charged chiral boson, above one day of the
data-taking is required.

\begin{figure}[th]
\epsfig{file=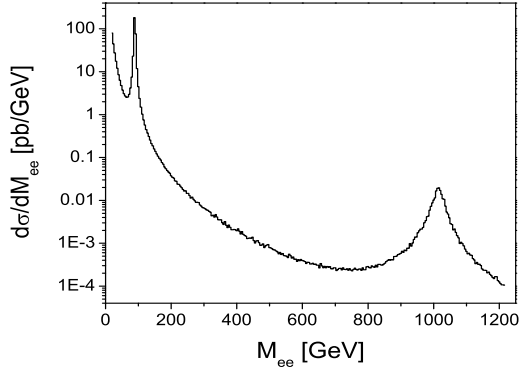,height=6cm,width=8cm} \caption{\label{fig:7} The $Z^0$
and $T^0$ boson peaks in the differential cross-section as a function of the
invariant dilepton mass. }
\end{figure}

However, the bad understanding of the detector resolution at the first runs and
the uncertainties in the transverse momentum of the heavy bosons smear the
Jacobian peak and the production of the chiral bosons could not be
distinguished from the production of the gauge bosons. More data and detailed
simulations will be needed. Therefore, the first crucial test of the discussed
model at the LHC will be the observation of the peak at 1 TeV in the Drell--Yan
dilepton channel (Fig.~\ref{fig:7}).

As it was discussed early, the shape of the peak as a function of the invariant
dilepton mass is the same as for the gauge and chiral bosons. The only
difference should be in the angular distributions of the lepton pairs, in
particular, their $p_T$ distributions. Instead the Jacobian peak at $p_T\simeq
M_V/2$ for the gauge bosons, a wide bump, well below this point, is expected
for the chiral boson in the lepton transverse distribution (see
Fig.~\ref{fig:2}).

\section*{Acknowledgements}

I am grateful to D. Kirilova for the overall help. I would like especially 
to thank A. Belyaev for introducing me into the CalcHEP package and 
continues assistance. 
I also appreciate the discussions with I. Boyko, C. Hof, H. Kim, J. Kim, 
V. Kim, L. Litov, C. Magass, A. Meyer, A. Pukhov, N. Skachkov, R. Tsenov.

\pagebreak[3]

\begin{figure}[th]
\includegraphics[angle=90,width=17.7cm]{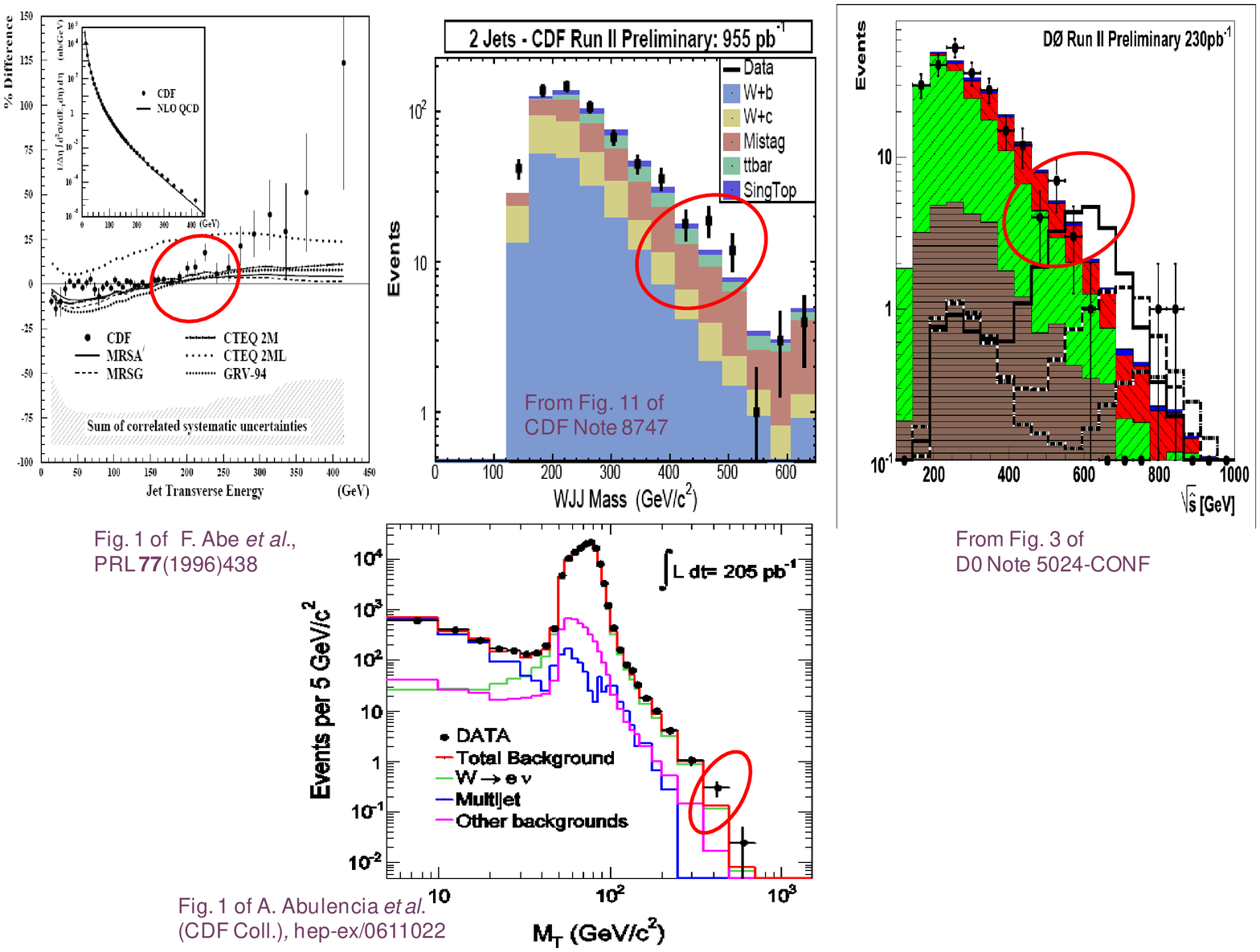}
\end{figure}

\end{document}